\documentclass[lettersize,journal]{IEEEtran}
\usepackage{amsmath,amsfonts}
\usepackage{algorithmic}
\usepackage{array}
\usepackage[caption=false,font=normalsize,labelfont=sf,textfont=sf]{subfig}
\usepackage{textcomp}
\usepackage{stfloats}
\usepackage{url}
\usepackage{verbatim}
\usepackage{graphicx}
\DeclareMathOperator*{\argmin}{arg\,min}
\usepackage{tabularx}
\newcolumntype{Y}{>{\centering\arraybackslash}X}
\usepackage{booktabs}
\def\BibTeX{{\rm B\kern-.05em{\sc i\kern-.025em b}\kern-.08em
    T\kern-.1667em\lower.7ex\hbox{E}\kern-.125emX}}
\usepackage{balance}
\begin{document}
\title{Joint graphical model estimation using Stein-type shrinkage for fast large scale network inference in scRNAseq data}
\author{Duong H.T. Vo, Nelofer Syed and Thomas Thorne
\thanks{Duong H.T. Vo and Thomas Thorne are with the Computer Science Research Centre, University of Surrey, United Kingdom. Nelofer Syed is with the John Fulcher Neuro-Oncology Laboratory, Department of Brain Sciences, Faculty of Medicine, Imperial College London, London W12 0NN, UK. Corresponding author email: tom.thorne@surrey.ac.uk}}

\markboth{Journal of \LaTeX\ Class Files,~Vol.~18, No.~9, September~2020}%
{Joint graphical model estimation using Stein-type shrinkage for fast large scale network inference in scRNAseq data}

\maketitle

\begin{abstract}
Graphical modeling is a widely used tool for analyzing conditional dependencies between variables and traditional methods may struggle to capture shared and distinct structures in multi-group or multi-condition settings. Joint graphical modeling (JGM) extends this framework by simultaneously estimating network structures across multiple related datasets, allowing for a deeper understanding of commonalities and differences. This capability is particularly valuable in fields such as genomics and neuroscience, where identifying variations in network topology can provide critical biological insights. Existing JGM methodologies largely fall into two categories: regularization-based approaches, which introduce additional penalties to enforce structured sparsity, and Bayesian frameworks, which incorporate prior knowledge to improve network inference. In this study, we explore an alternative method based on two-target linear covariance matrix shrinkage. Formula for optimal shrinkage intensities is proposed which leads to the development of JointStein framework. Performance of JointStein framework is proposed through simulation benchmarking which demonstrates its effectiveness for large-scale single-cell RNA sequencing (scRNA-seq) data analysis. Finally, we apply our approach to glioblastoma scRNA-seq data from \cite{yeo2022single}, uncovering dynamic shifts in T cell network structures across disease progression stages. The result highlights potential of JointStein framework in extracting biologically meaningful insights from high-dimensional data.
\end{abstract}


\section{Introduction}
The introduction of single-cell RNA sequencing (scRNAseq) using high-throughput next-generation DNA sequencing (NGS) transformed transcriptomic research by providing data at a higher resolution with information on cellular states and the molecular interactions of individual cells \cite{ziegenhain2017comparative}. The development of scRNAseq has allowed researchers to collect transcriptomic information from more than 20,000 genes from more than 10,000 cells in one experiment \cite{jovic2022single}. This requires computational methods that support big data analysis while maintaining performance in inference processes such as network estimation.

Gene network inference is one of the active research fields where computational biologists aim to extract interaction information between genes or their products. These networks can include DNA-protein interaction, protein-protein interaction or gene regulatory network. In the studies Pearson or Spearman correlation tests are often used to build correlation matrices which contribute to the construction of gene co-expression networks. However, these tests fail to exclude indirect correlations between variables under third-party effects \cite{huang2010learning}. Graphical modeling instead records conditional independencies between random variables in a connectivity graph \cite{giraud2021introduction}. To extract direct connections between variables in graphical modeling, the partial correlation matrix is often used. Multiple methods have been developed for graphical modeling under the assumption of high sparsity in the final network \cite{schafer2005shrinkage, friedman2008sparse, meinshausen2006high}. 

Joint graphical modeling, an extension to standard graphical models, collates a common network structure from other related networks to improve the overall performance of network inference compared to separate estimation \cite{tsai2022joint}. Joint graphical modeling acts on multiple groups of observations for a common set of variables, learning a specific network for each set of observations. In the case of biological network inference, joint network estimation supports the identification of conserved networks between different biological groups, whilst also highlighting specific edges in each group. Its application in biology has been highlighted in a protein signaling pathway study \cite{oates2014joint}. Multiple approaches have been developed for modeling joint network structures, including regularization and Bayesian methods. In this study, the application of two-target linear covariance matrix shrinkage (TTLS) as a new approach to joint graphical modeling is explored. TTLS method extends Stein-type linear shrinkage, an approach in standard graphical modeling that does not incorporate joint estimation. Instead, it separately estimates the covariance matrix for each group which ensures independent shrinkage within distinct subsets of data. Stein-type covariance shrinkage has been widely use to shrink the sample covariance matrix towards the identity matrix \cite{ledoit2003honey}. In TTLS, rather than shrinking the sample covariance to the identity matrix, the sample covariance matrix is shrunk towards a shared covariance matrix containing information on the common network structure shared between all groups. 

With the rapid pace of development in single-cell RNA sequencing technology, inference methods able to perform joint graphical modeling on large-scale datasets are becoming increasingly important. In this study, TTLS is implemented in a framework we refer to as \textbf{JointStein}, which is designed for large-scale joint graphical model inference. JointStein is benchmarked against current joint graphical modeling approaches in terms of performance and computational time. We also apply JointStein to joint network inference in experimental data from glioblastoma and \textit{Plasmodium falciparum} scRNAseq data \cite{yeo2022single, poran2017single}.

\section{Methods}
\subsection{Overview of joint graphical model estimation}
Graphical modeling allows us to extract conditional dependence and independence relationships between random variables \cite{giraud2021introduction}. Conditional independency is defined as the case where the joint probability of two variables $X$ and $Y$ conditional on a third variable $Z$ can be rewritten as:
\begin{equation}
    P(X,Y|Z) = P(X|Z)P(Y|Z)
\end{equation}

This states that given knowledge of $Z$, the two variables $X$ and $Y$ are independent of one another.

Gaussian Graphical Models (GGMs) are used to describe the conditional dependence structure among multiple Gaussian-distributed variables \cite{giraud2021introduction}. In these models, edges between nodes represent direct interactions, as captured by the nonzero elements of the precision matrix, after adjusting for the influence of all other variables \cite{giraud2021introduction}. One popular approach for Gaussian graphical modeling is through estimating the partial correlation matrix by inverting the unknown and nonsingular population covariance matrix $\Sigma$ \cite{giraud2021introduction}. The population covariance matrix $\Sigma$ is often estimated from the sample covariance matrix $S$. The implementation of graphical models for real-world network inference has been proposed in various fields such as biology, economy and finance \cite{giraud2021introduction}, and in most cases, the underlying network is assumed to be sparse. To induce sparsity in the network and overcome the large $p$ small $n$ challenge, in which sample covariance matrix S becomes singular, various different methods of graphical model inference have been developed. These include minimizing the negative log-likelihood with $L_1$ regularization, and shrinking the sample covariance matrix $S$ towards the identity matrix \cite{friedman2008sparse, meinshausen2006high, ledoit2004well}. \par

\begin{figure}[!t]
    \centering
    \includegraphics[width=0.7\linewidth]{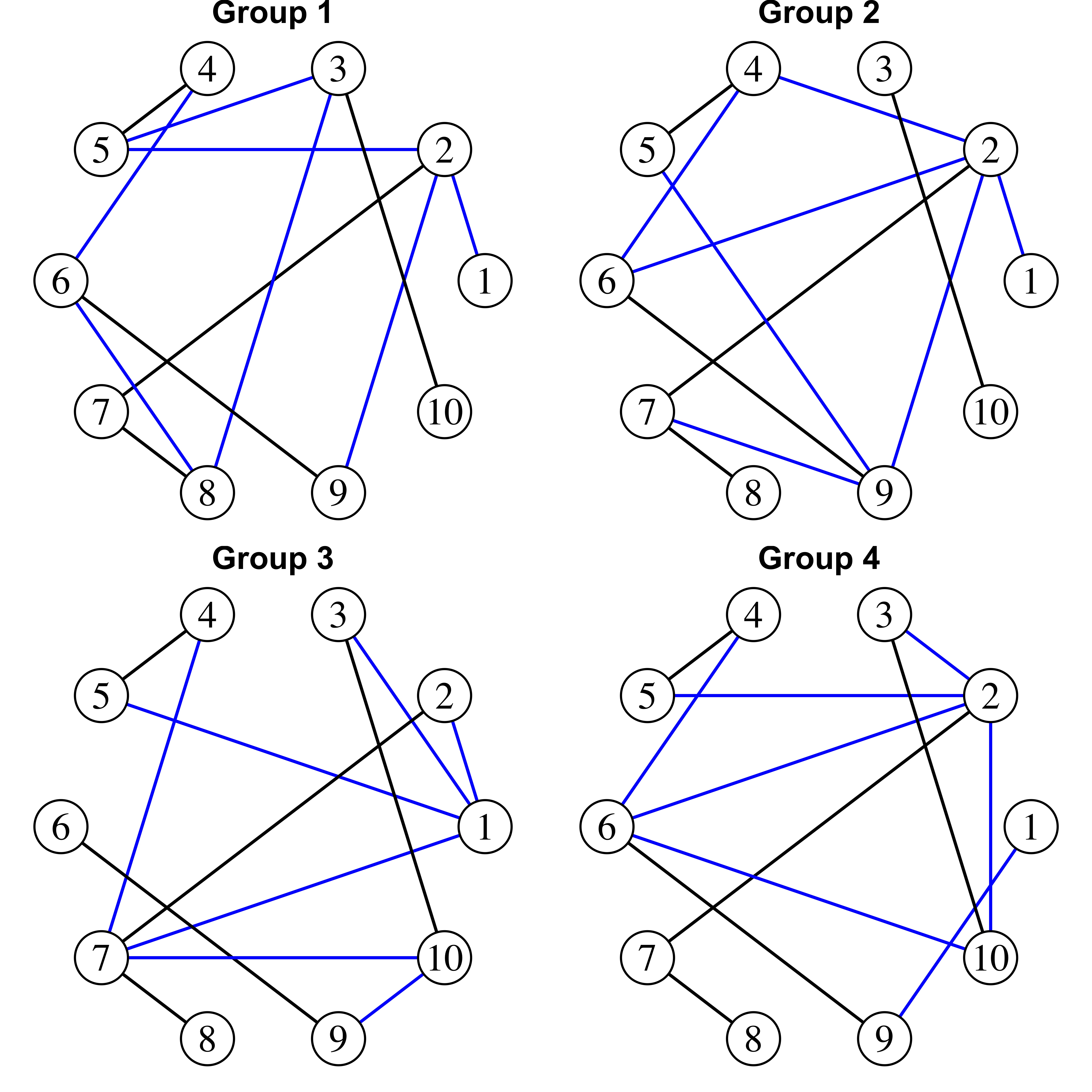}
    \caption{\textbf{Joint estimation of related networks.} The shared network between groups is highlighted as black edges, whilst unique edges to each group are coloured in blue.}
    \label{fig:jointGraph}
\end{figure}

An extension to standard graphical modeling is the application of joint graphical modeling, as the presence of a partially shared network structure, allowing information sharing between related data sets, has been shown to boost the power of network inference, especially in high-dimensional data analysis \cite{tsai2022joint}. The idea of joint graphical modeling is illustrated in Figure \ref{fig:jointGraph}. Specifically, in heterogeneous data when multiple related groups are present, joint estimation aims to borrow and extract a common network structure between all of the groups. Then individual graphs containing some of the shared structure, but also unique group specific edges can be inferred. This allows us to identify network edges that are shared across multiple groups, while also allowing distinct connections in individual groups. In the case of gene network inference, joint estimation can reveal both core and discrete genetic network structures in different environments or cell types \cite{guo2011joint}.

To implement joint graphical model in real-world data, there are some considerations. Firstly, when the number of features exceeds the number of samples, the sample covariance matrix becomes singular and noisy \cite{ledoit2003honey}. This same challenge is also faced in standard graphical modeling. With advances in single-cell sequencing technology, gene expression data can be collected from many thousands of genes and potentially millions of cells \cite{jovic2022single}. Hence, network inference methods for large-scale applications are essential to enhance computational analysis with less computational resource requirements, for instance, methods with faster computing time. Lastly, for joint estimation, approaches to borrow and incorporate shared network information in network inference process are of paramount importance, especially in a biological context where there is likely to be some shared network structure between different groups, for example between different cell types or experimental conditions.

In recent works, there are two main approaches for joint graphical modeling, regularization-based and Bayesian methods \cite{tsai2022joint}. Similar to the graphical lasso, sparsity in regularization-based approaches is induced by $L_1$ regularization on entries of the inverse covariance matrix, also referred to as the precision matrix \cite{mazumder2012graphical}:   

\begin{equation}
    \hat{\Theta} = \underset{\Theta \in \mathbb{R}^{p \times p}}{argmax} \{log(det(\Theta)) - tr(S\Theta) - \lambda_1\parallel\Theta\parallel_1\}
\end{equation}

In regularization-based joint graphical modeling, a second penalty function is often added to the estimator which acts as similarity constraint between groups \cite{tsai2022joint}. For example, the Fused Graphical Lasso (FGL) and Group Graphical Lasso (GGL) add $P_{FGL}(\Omega) = \lambda_2\sum_{k < k'}\sum_{i,j}|w_{i,j}^{(k)} - w_{i,j}^{(k')}|$ and  $P_{GGL}(\Omega) = \lambda_2\sum_{i \neq j}\{\sum_{k=1}^K(w_{i,j}^{(k})^2\}^\frac{1}{2}$ to their objective functions, respectively \cite{danaher2014joint}. Both FGL and GGL are under the broader joint graphical lasso (JGL) framework which is introduced by \cite{danaher2014joint}.

These added functions encourage some shared network pattern between groups while the conventional penalty function ensures sparsity induction on final graphical model. However, in implementation, selecting appropriate values of the regularization parameters and the computational time required by the algorithms are two main hindrances.

On the other hand, Bayesian models encode the properties of the group of graphical models in their prior. For instance, \cite{jia2021fast} designed priors specific to temporal and spatial models, encouraging edges to be shared between networks that are temporally or spatially close to one another. However this algorithm is complex to implement and can be computationally demanding, especially when using Markov Chain Monte Carlo algorithms to estimate the posterior distribution.

In this study, we aim to explore a different approach of joint graphical modeling using Stein-type shrinkage with two target matrices, and benchmark the performance of two-target Stein-type linear covariance matrix shrinkage through performance comparison analysis on simulated data. 

\subsection{Two-target linear covariance matrix shrinkage}
A new approach to joint graphical modeling we explore in this work is called two-target linear shrinkage (TTLS). The Two-target Linear Shrinkage (TTLS) formula is based on the double shrinkage estimator from \cite{ikeda2016comparison}, and the multi-target shrinkage estimation from \cite{lancewicki2014multi}. To combine information of the common network structure while shrinking entries of sample covariance matrix towards zero in our approach, we apply TTLS balancing between a shared covariance matrix $S_{shared}$ and the identity matrix $I$ using shrinkage regulators $\gamma_1$ and $\gamma_2$ as follows:

\begin{equation}
    \Sigma = (1 - \gamma_1 - \gamma_2)S + \gamma_1 S_{shared}^{(i)} + \gamma_2I
\end{equation}

Then the estimator in TTLS is as follow, for $i \in G$ groups:
\begin{equation}
    \hat{\Sigma}_i = (1-\hat{\gamma_1}-\hat{\gamma_2})S_i + \hat{\gamma_1} \hat{S}_{shared}^{(i)} + \hat{\gamma_2}I
\end{equation}

The shared covariance matrix $S_{shared}$ contains information of the underlying common network structure in our TTLS approach. For an estimator of $S_{shared}$, we use the average of the sample or shrunk covariance matrices of the other groups can be used. Shrinking the sample covariance matrix before taking the average can potentially reduce errors and ensure the estimator is non-singular in case of high-dimensional data:
\begin{equation}
    \hat{S}_{shared}^{(i)} = \frac{1}{G-1}\sum_{j\neq i}^G S_j
\end{equation}

The risk function of $\hat{\Sigma_i}$ is based on the Frobenius loss:

\begin{equation}
    \begin{aligned}
        R(\hat{\Sigma}_i, \Sigma_i) & = E\{\|\hat{\Sigma}_i - \Sigma_i\|^2_F\} \\
        & = E\{\|S_i-\Sigma_i\|^2_F\} + \gamma^TA\gamma - 2\gamma^Tb
    \end{aligned}
\end{equation}

where

\begin{equation}
    \gamma = 
    \begin{pmatrix}
        \hat{\gamma_1} \\
        \hat{\gamma_2}
    \end{pmatrix}
\end{equation}

\begin{small}
    \begin{equation}
    A = 
        \begin{pmatrix}
            E\{\|\hat{S}_{shared}^{(i)} - S_i\|^2_F\} & E\{\langle \hat{S}_{shared}^{(i)} - S_i, I - S_i \rangle\} \\
            E\{\langle I - S_i, \hat{S}_{shared}^{(i)} - S_i \rangle\} & E\{\|I - S_i\|^2_F\}
        \end{pmatrix}
    \end{equation}
\end{small}

\begin{equation}
    b = 
    \begin{pmatrix}
        E\{\langle \hat{S}_{shared}^{(i)} - S_i, \Sigma - S_i \rangle\} \\
        E\{\langle I - S_i, \Sigma - S_i \rangle\}
    \end{pmatrix}
\end{equation}

From \cite{fisher2011improved}, $E\{\|S_i-\Sigma_i\|^2_F\}$ equals $\frac{1}{n}a_2 + \frac{p}{n}a^2_1$ where $a_i=\frac{1}{p}tr(\Sigma^i)$ and details of the calculation of $a_1$ and $a_2$ are in \cite{fisher2011improved}. Most importantly, value of $E\{\|S_i-\Sigma_i\|^2_F\}$ is shown not to depend on $\gamma$. Therefore, the optimization problem can be rewritten into:
\begin{equation}
    \argmin F(\gamma) = \gamma^TA\gamma - 2\gamma^Tb 
\end{equation}

with 
\[ \hat{\gamma_1} \geq 0 \]
\[ \hat{\gamma_2} \geq 0 \]
\[ \hat{\gamma_1} + \hat{\gamma_2} \leq 1. \]

Estimation of the matrix A uses its sample based counterparts and the vector b is based on equation 30 in \cite{lancewicki2014multi}:
\begin{equation}
    A = 
    \begin{pmatrix}
        \|\Bar{S}_{shared}^{(i)} - S_i\|^2_F & \langle \Bar{S}_{shared}^{(i)} - S_i, I - S_i \rangle \\
        \langle I - S_i, \Bar{S}_{shared}^{(i)} - S_i \rangle & \|I - S_i\|^2_F
    \end{pmatrix}
\end{equation}

\begin{equation}
    b = 
    \begin{pmatrix}
        \hat{V}(S_i) -\hat{V}(\hat{S}_{shared}^{(i)}) \\
        \hat{V}(S_i) -\hat{V}(I)
    \end{pmatrix}
\end{equation}

Calculations of $\hat{V}(S_i)$, $\hat{V}(\hat{S}_{shared}^{(i)})$ and $\hat{V}(I)$ are as following, considering a squared symmetric matrix M:
\begin{equation}
    \hat{V}(M) = \frac{n}{(n-1)^2(n-2)}\sum_{i=1}^n\|(M_i - \frac{n-1}{n}M)\|^2_F
\end{equation}

To solve the optimization problem in equation 10, we use Lagrange multipliers with slack variables to calculate the formula of the optimal shrinkage intensity $\gamma_1$ and $\gamma_2$. Details of the calculation are in Supplementary material 1. Another option for solving objective functions is using Nelder-Mead optimization algorithm with inequality constraints provided by \texttt{constrOptim} function in R.

\subsection{Higher criticism in significance testing of sparse partial correlation coefficients when number of features is high}
Higher criticism (HC) is defined as a second-level significance test used when a very small fraction of the independent tests are not in the null hypothesis \cite{donoho2004higher}. This is widely applicable in the case where signals are sparse and using standard approaches for significance testing such as a $0.05$ threshold on adjusted p-values cannot decrease the false positive rate.

In the case of estimating sparse partial correlation matrices in graphical modeling, significance testing is required to select significantly non-zero entries in the matrix. This is often done using t-statistics in a Pearson correlation test of significance. However, when the number of features (p) increases, the number of coefficients $\frac{1}{2}p(p-1)$ for testing grows quadratically. If on average there are $2$ edges per node, the percentage of nonzero entries in the population partial correlation matrix is then $\frac{4}{p-1}$, and as p goes towards infinity, the fraction is extremely small and goes towards zero. Hence, a second-level of testing to reduce false positive rates is necessary and important in this sparse signal case.

Higher criticism was first generalized by \cite{donoho2004higher} in which it is a test on increasingly-ordered p-values. The objective function of HC statistics proposed by \cite{donoho2004higher} is:
\begin{equation}
    HC = max_{1\leq i\leq \alpha_0N}\sqrt{N}[i/N - p_{(i)}]/\sqrt{p_{(i)}(1-p_{(i)})}.
\end{equation}
From the first generalized function of HC statistics, multiple objective functions have been proposed which aim to solve different problems such as $p > n$ high-dimensional data \cite{donoho2009feature}. In this study, two HC objective functions from \cite{donoho2009feature} ($HC_{DJ}$) and \cite{li2015higher} ($HC_{LS}$) are implemented on false discovery rate (FDR) adjusted p values. Denoting $p_{(i)}$ is the i-th p-value in list of p-values arranging in ascending order, $n$ is total number of p-values or hypothesis tests in the list:
\begin{equation}
    HC_{DJ} = \max_{1\leq i \leq \alpha_0} \sqrt{n}[i/n - p_{(i)}]/\sqrt{p_{(i)}(1-p_{(i)})}
\end{equation}

\begin{equation}
    \begin{aligned}
         HC_{LS} = & \max_{k_0 \leq k \leq k_1}\sqrt{2n}I\{p_{(k)} < k/n\} \\
          & \sqrt{(k/n)log(k/np_{(k)}) - (k/n-p_{(k)})]} \\
    \end{aligned}
\end{equation}

The HC formula determines a position $t$ in the list of ordered p-values, giving a cut-off point at which all p-values with positions $i \leq t$ are considered to belong to the alternative hypothesis. Partial correlation coefficients with HC-significant p-values are considered to be non-zero.

\section{Results and Discussion}
\subsection{JointStein workflow and benchmarking process}
The JointStein approach is a joint graphical modeling workflow based on Stein-type linear covariance matrix shrinkage. The workflow comprises three main steps, data processing, two-target linear shrinkage and significance testing. In the first step, when working with scRNA-seq data, heterogeneous count data containing samples from multiple unknown groups can be classified into discrete clusters using existing clustering methods \cite{van2008visualizing, mcinnes2018umap}. However, in some cases when groups or cell types are known, and this clustering step can be skipped and normalization can be performed on the count data. After the data processing step, two-target linear covariance matrix shrinkage is applied which produces an estimated partial correlation matrix. Standard significance tests with p-value adjustment are then applied, and when the number of features is above 1000, higher criticism is recommended for estimating significant partial correlation coefficients.

\begin{figure}[!t]
    \centering
    \includegraphics[width=0.7\linewidth]{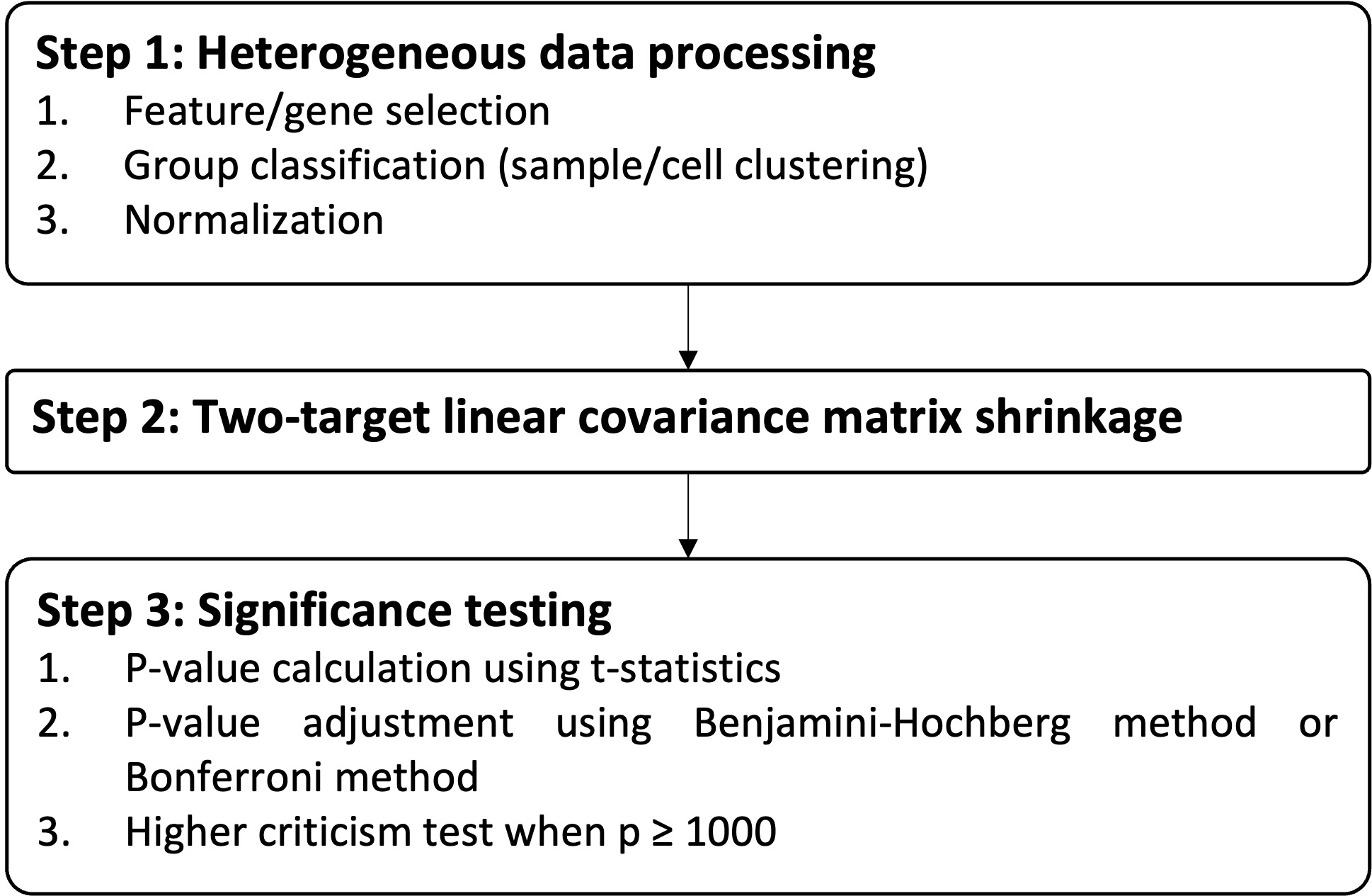}
    \caption{\textbf{JointStein workflow.} JointStein is a Stein-type joint estimation framework which implements two-target linear covariance matrix shrinkage and higher criticism when the number of features is equal to or larger than 1000.}
    \label{fig:JSworkflow}
\end{figure}

The JointStein workflow is benchmarked against standard linear shrinkage (GeneNet \cite{schafer2005shrinkage}), regularization-based joint estimation (JGL \cite{danaher2014joint}) and Bayesian based joint estimation (Bayes \cite{jia2021fast}). The benchmarking process is carried out on simulated data, and experimental single-cell RNA sequencing (scRNAseq) data. Simulation benchmarking is conducted by simulating data from a multivariate normal distribution, with precision matrices simulated using the model from Figure \ref{fig:jointGraph}, in which each group comprises of 40\% shared edges and 60\% randomly generated edges. Matthew's correlation coefficient (MCC), which is a correlation coefficient between a set of predicted and reference values, is chosen as the performance metric. Further details of the benchmarking are in provided in Supplementary material 2.

To demonstrate the applicability of the method to experimental data, scRNAseq data from classified brain cells during tumor progression and from malaria parasite (\textit{Plasmodium falciparum}) at different development stages are used to demonstrate the potential of joint network estimation in real-world data \cite{yeo2022single,poran2017single}. 

\subsection{Performance comparison to existing joint graph estimation approaches}

\begin{figure}[!t]
    \centering
    \includegraphics[width=\linewidth]{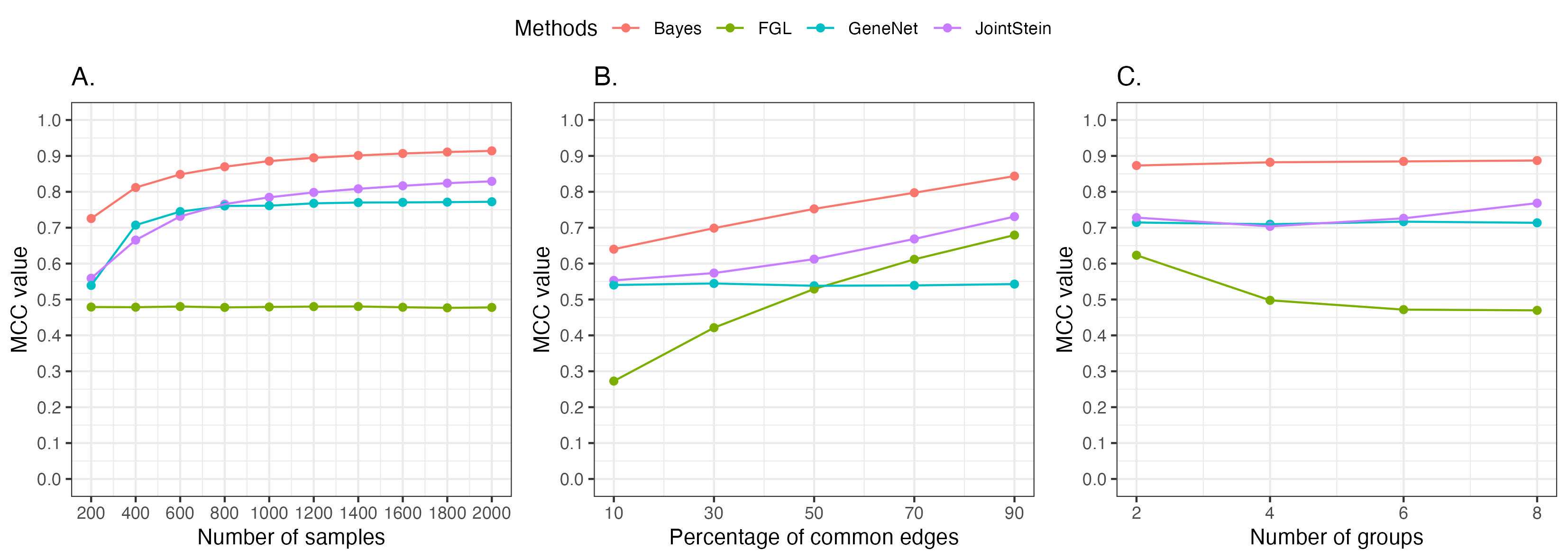}
    \caption{\textbf{Performance of JointStein against graphical modeling existing methods in multivariate normally distributed data.} JointStein framework is compared when there is a variation in number of samples ($p=2000$, $G=50$, proportion of common edges is 40\%) (A), percentage of common edges ($p=2000$, $n=200$, $G=5$) (B) and number of groups ($p=2000$, $n=1000$, proportion of common edges is 40\%)(C) with 10 iterations.}
    \label{fig:performComp}
\end{figure}

Simulation benchmarking was conducted using multivariate normally distributed data with joint network simulation. The performance assessment was conducted under various conditions by adjusting the sample size, the proportion of shared edges among networks, and the number of groups. To examine the effect of sample size, the number of observations ($n$) was varied between $200$ and $2000$, while the number of features ($p$) remained constant at $2000$ across 5 groups ($G$), with 40\% of edges being shared across networks. Furthermore, the impact of shared edge proportion was analyzed by fixing $p=2000$, $n=200$, and $G=5$, while modifying the proportion of shared edges from 10\% to 90\%. Lastly, the influence of the number of groups was evaluated by adjusting $G$ from 2 to 8, keeping $p=2000$, $n=1000$, and maintaining 40\% of edges as shared across networks. The results are summarized in Figure \ref{fig:performComp}. In all cases, the Bayesian method outperforms other methods, with the JointStein approach second. Noticeably, $l1$ regularization based joint estimation, fused graphical lasso algorithm (FGL) does not increase in performance as number of samples rises.

A potential issue explaining the performance of FGL is the fixed regularization parameters ($\lambda_1=0.3$ and $\lambda_2=0.1$), as current regularization parameter selection methods are developed for standard graphical modeling \cite{zhao2012huge} and are not applicable to joint approaches. Hence, choosing the optimal regularization parameters for FGL remains a subject for research. Compared to FGL, JointStein is nonparametetric, in which shrinkage intensities are formulated from minimizing objective functions. On the other hand, GeneNet performance remains constant when the number of groups or percentage of common edges increases. This is to be expected as the method is designed for inference in a single group and does not exploit the shared network structure present in the data. This simulation benchmarking illustrates the potential of incorporating shared-network information to improve the performance of graphical modeling.

To explore the practicality of joint graphical modeling approaches in large-scale data analysis, the computational time of Bayes, FGL, JointStein and the GeneNet approach are measured and highlighted in Table \ref{tab:CompTime}. JointStein has by far the shortest computational time among the two other joint estimation methods (Bayes and FGL). Importantly, in large count matrix data (p=2000, n=1000, G=5), it takes approximately $7$ and $6$ hours for Bayes and FGL to complete the estimation, respectively whilst, it takes under $3$ minutes for JointStein. This emphasizes potential of JointStein to conduct joint estimation analysis when there is a large number of genes. Importantly, the advantage of JointStein becomes more pronounced as the number of cell groups increases. For instance, in \cite{yeo2022single}, 36 distinct cell clusters were annotated from glioblastoma samples, and joint graphical modeling methods like Bayes and FGL would be computationally prohibitive in such large-scale multi-group settings. In contrast, the faster computation of JointStein suggests its suitability for modeling gene networks across numerous cell types, making it a promising tool for large-scale single-cell transcriptomics analysis.

\begin{table}
    \renewcommand{\arraystretch}{1.2}
    \caption{\textbf{Computational time.} All analyses were performed on Ubuntu with Intel Xeon Gold 5215L CPU (10 iterations, p - number of features/genes, n - number of observations/cells in each group, G - number of groups, s - seconds, m - minutes, h - hours).}
    \begin{tabularx}{\linewidth}{ccccccc}
        \toprule
        p & n & G & {Bayes} & {FGL} & {GeneNet} & {JointStein} \\
        \midrule
        200 & 100 & 5 & 6.17(m) & 2.09(m) & 0.03(s) & 0.17(s) \\
        1000 & 500 & 5 & 52.86(m) & 1.17(h) & 3.220(s) & 16.742(s) \\
        2000 & 1000 & 5 & 7.02(h) & 6.06(h) & 29.722(s) & 2.48(m) \\
        \bottomrule
    \end{tabularx}
    \label{tab:CompTime}
\end{table}

\subsection{Implementation in experimental scRNAseq data}

Visualization and interpretation of gene expression data in scRNAseq experiments are of paramount importance. Multiple methods are proposed including heatmaps and clustering to identify biologically related genes, gene set enrichment analysis (GSEA) for functional annotation and network analysis to demonstrate how different pathways interact \cite{EMBL}. In this experimental implementation, we aim to use the JointStein workflow for network inference, while combining the network based results with heatmaps, clustering, and GSEA, to explore the scRNAseq data from different angles. First, gene expression data of scRNAseq studies after quality control and normalization are collected. Cell clustering and cluster identification then allow distinct groups to be identified for use in our joint estimation approach. The JointStein workflow is then applied to the scRNAseq data of each group to infer gene networks, and following this simplified networks using gene clusters with their corresponding annotated functions as nodes are used to analyze interactions between different pathways.

\begin{figure}[!t]
    \centering
    \includegraphics[width=\linewidth]{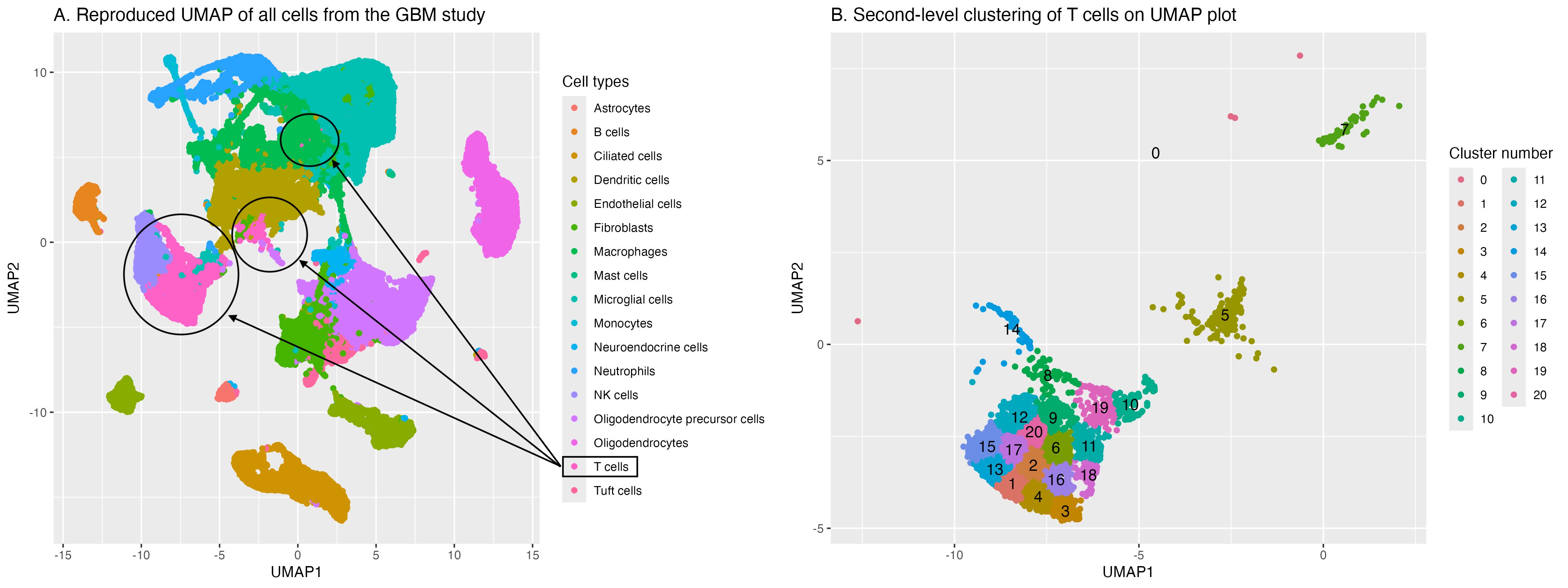}
    \caption{\textbf{Cell cluster plots of glioblastoma study \cite{yeo2022single}.} A, UMAP plot of all cells with cell type annotation reproduced from the GBM study \cite{yeo2022single}. B, Second-level clustering to identify subclusters of T cells.}
    \label{fig:GBMUMAP}
\end{figure}

Glioblastoma (GBM) is malignant brain cancer with a highly immunosuppressive and protumorigenic immune microenvironment \cite{yeo2022single}. SCRNAseq technology was used to characterise changes in the immune landscape during glioblastoma progression in a \textit{de novo} orthotopic mouse model \cite{yeo2022single}. GBM single cells were harvested at early, late stages of GBM and classified by cell markers \cite{yeo2022single}. Normal brains were included as controls \cite{yeo2022single}. The cell clustering UMAP plot was reproduced from the study (Figure \ref{fig:GBMUMAP}A). In this study, we focused on T cells within the scRNAseq dataset to demonstrate the relevance of JointStein framework in joint network inference for single-cell data analysis. T cells play a critical role in the immune response to GBM and have been extensively studied in cancer research, providing a well-characterized reference for evaluating network inference results \cite{wang2021different}. Their biological, genomic, and gene regulatory data in GBM are well-documented, allowing for direct comparison with existing literature. This enables us to assess the consistency of our inferred networks with known gene interactions, further validating the robustness of the proposed framework. 

Further stratification was carried out since T cells are shown to consist of multiple subpopulations, each with distinct biological functions and roles in immune responses \cite{koh2023cd8}. The original GBM study annotated this cluster broadly as T cells without distinguishing finer subpopulations. Further clustering was applied using the kNN algorithm to refine the classification and identify subclusters that may better represent biologically distinct T cell populations. The final clustering result identified 20 subclusters of T cells and is presented in Figure \ref{fig:GBMUMAP}B. 

\begin{figure*}[!t]
    \centering
    \includegraphics[width=\linewidth]{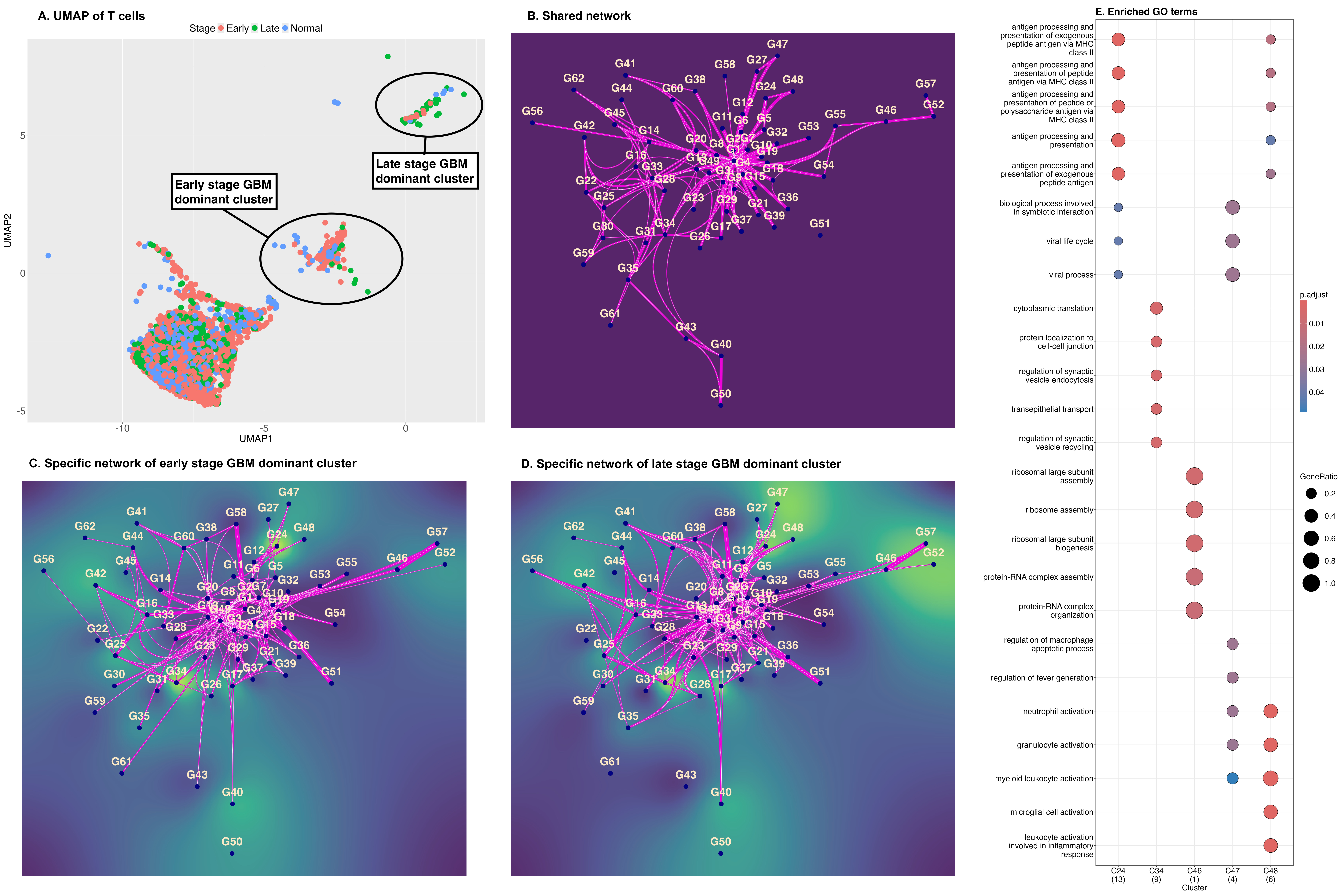}
    \caption{\textbf{Joint gene network inference of glioblastoma study \cite{yeo2022single}.} A, UMAP of harvested T cells from 3 different stages were reproduced from main study and smaller clusters were classified using kNN algorithm. The result is in Figure \ref{fig:GBMUMAP}B. Cluster 5 and 7 forming clear distinct clusters with high percentages of cells from early and late stages of GBM are subjects of interest. B, Shared network is shown between all T cell clusters in the UMAP with thickness corresponding amount of gene interactions between clusters. C \& D, Specific networks to early-GBM and late-GBM clusters, respectively are illustrated with average gene expression in each clusters as heatmap background. }
    \label{fig:GBMNet}
\end{figure*}

In this implementation, scRNAseq data of T cells from three stages and after data quality control is used. Results are shown in Figure \ref{fig:GBMNet}. In this analysis, a total of 3,381 differentially expressed genes were included to construct gene networks. To facilitate visualization and interpretation, a gene clustering step was performed with the objective of reducing the number of nodes and dimensions, enabling a clearer representation of gene relationships while integrating gene set annotations that reflect biological functions. Genes were clustered based on their expression profiles in T cells, allowing functionally related genes to be grouped together. Hierarchical clustering was employed to identify gene clusters, where Euclidean distance was used to compute the distance matrix, and the complete agglomeration method was applied for hierarchical clustering. To refine the clustering, we implemented the hybrid dynamic tree-cutting method using the \texttt{cutreeDynamic} function in R, with the \texttt{deepSplit} parameter set to 4, ensuring a minimum of four genes per cluster. This approach resulted in the identification of 62 gene sets which were annotated functionally using PANTHER overrepresentation enrichment analysis \cite{mi2017panther}. Final results of gene sets and their annotations were subsequently integrated into the network plots to enhance biological interpretation.

Noticeably, clusters 5 and 7 have high proportions of early and late stages of GBM cells respectively. Hence, these clusters are of interest for joint network inference and annotated as early-GBM and late-GBM clusters. The shared network structure among all clusters is visualized in Figure \ref{fig:GBMNet}B while the specific networks of early-GBM and late-GBM clusters are in Figure \ref{fig:GBMNet}C and D, respectively. From the network inference results, gene sets 24 and 34 are highly expressed in both of these clusters. Gene set 24 comprises genes primarily involved in antigen processing and immune response. This includes C1qa, C1qb, and C1qc, which encode complement component C1q, a key protein in both adaptive and innate immunity \cite{reid2018complement}. C1q recognizes and binds to various immune system activators, playing a crucial role in initiating complement activation, clearing cellular debris, and regulating immune cell interactions \cite{reid2018complement}. Structurally, it consists of six identical subunits forming a collagenous stalk with globular heads, which facilitate its binding functions \cite{MIDDLETON2016318}. Additionally, this cluster contains H2-Aa, H2-Ab1, and H2-Eb1, encoding major histocompatibility complex (MHC) class II proteins. These proteins are essential for presenting processed antigens to T cells, a critical step in triggering antigen-specific immune responses \cite{holling2004function}. Beyond their role in antigen presentation, MHC class II molecules are also involved in intracellular signaling pathways that can lead to apoptosis \cite{holling2004function}. Furthermore, Tuba1b and Tubb5 are present in this cluster, encoding tubulin proteins that provide structural integrity to the cell, supporting cytoskeletal organization and stability \cite{binarova2019tubulin}. Cluster 34 is enriched with genes encoding actin proteins, including Actg1 and Actb. Actin filaments serve multiple essential functions, including providing mechanical stability, facilitating intracellular transport, and enabling cell movement \cite{pollard2009actin}. 

In late-GBM cell cluster, gene set 46, 47, 48, 52 and 57 are highly activated. Gene set 46 contains genes involved in mitochondrial respiration, including mt-Nd1, mt-Nd4, mt-Cytb, and mt-Co3. mt-Nd1 and mt-Nd4 encode subunits of NADH dehydrogenase, an essential enzyme in the electron transport chain responsible for transferring electrons from NADH to ubiquinone, contributing to ATP production \cite{yagi2006possibility}. mt-Cytb encodes cytochrome B, a key component of respiratory chain complex \cite{everse2013heme}. mt-Co3 encodes cytochrome c oxidase, the terminal enzyme in the electron transport chain that catalyzes the reduction of oxygen to water, driving oxidative phosphorylation \cite{ishigami2023structural}. Together, these proteins play a fundamental role in mitochondrial metabolism by enabling efficient cellular respiration. Gene set 47 consists of Ccl5, Ctsb, and Ctsd, which are involved in the regulation of macrophage apoptosis \cite{novak2020ccr5, kranjc2019cytokine, ding2022cathepsins}. These genes have been implicated in immune system modulation and reported to contribute to the poor prognosis of glioblastoma (GBM) \cite{novak2020ccr5, kranjc2019cytokine, ding2022cathepsins}. Ccl5 (C-C motif chemokine ligand 5) plays a role in inflammatory responses and immune cell recruitment, while Ctsb and Ctsd encode cathepsins, a lysosomal protease that regulate protein degradation and apoptosis \cite{marques2013targeting, dheer2019cathepsin}. Their involvement in GBM progression highlights their potential role in tumor-associated inflammation and immune evasion. Cluster 48 includes Lyz1 and Lyz2, which encode lysozymes. Lysozymes play a crucial role in host defense by breaking down bacterial cell walls, contributing to antibacterial immunity \cite{ragland2017bacterial}. These genes have been reported to be particularly responsive to bacterial infections, underscoring their importance in pathogen defense and immune regulation \cite{ragland2017bacterial}. 

Edges between gene groups represent connections between individual genes belonging to each group. For example, the presence of an edge between group 24 and group 48 indicates that at least one gene in group 24 is linked to one or more genes in group 48. The thickness of each edge corresponds to the number of such connections, with thicker edges representing a higher number of gene-to-gene interactions between the two groups. Connection between group 24 and 48 is present in all networks of cell clusters while connection between group 24 and 47 is only in the networks of early-GBM and late-GBM cell clusters. In the context of the T cell gene network from glioblastoma samples, the interaction between clusters 24 and 48 may reflect coordinated immune responses within the tumor microenvironment. Gene cluster 24 includes genes involved in antigen processing and complement activation, such as C1qa, C1qb, and C1qc, which play a role in recognizing pathogens and modulating immune cell activity. Meanwhile, gene cluster 48 contains Lyz1 and Lyz2, encoding lysozymes that contribute to antibacterial defense. The presence of these clusters within the T cell network suggests a potential link between complement-mediated immune signaling and immune clearance responses. Given that complement activation can influence macrophage function and enhance phagocytosis, it is possible that interactions between these clusters contribute to shaping the immune landscape in glioblastoma by regulating immune clearance mechanisms and inflammatory responses \cite{bohlson2014complement}. This interaction may be particularly relevant in the context of glioblastoma, where immune modulation plays a critical role in shaping tumor progression and immune evasion. 

In both early-GBM and late-GBM networks, gene group 50, which consists of pseudogenes (Gm15464, Gm10689, Gm17786, Gm8618, Gm10054) in \textit{Mus musculus}, remains isolated, lacking connections to other gene groups. While little is known about the functions of these pseudogenes, their identification in differential gene expression analysis suggests a potential role in immune modulation within T cells in the glioblastoma microenvironment. In contrast, gene group 17 exhibits interactions exclusively in the early-GBM cluster, forming connections with gene groups 4, 9, and 39. Notably, the connection between gene group 17 and 9 includes edges between Tmsb10 and Calm1. Tmsb10 encodes a transcriptional factor crucial for cytoskeletal organization and has been associated with immune cell activity, including T cells (gamma delta, CD4$+$ Th1 and Th2, CD8$+$), macrophages (M1 and M2) \cite{xiao2019tmsb10, li2023pan}. Meanwhile, Calm1 encodes calmodulin, a calcium sensor involved in intracellular $Ca^{2+}$ signaling, playing a fundamental role in cellular processes and indirectly contributing to T-cell and B-cell activation \cite{Hasterok2022CALM1}. Gene group 17 connects to gene group 39 through edges between Cfl1 and Arhgdib. Cfl1 encodes the cofilin protein, which is essential for actin cytoskeletal remodeling. It has been demonstrated that cofilin plays a critical role in early $\alpha\beta$ T-cell development, while also exhibiting differential involvement in $\alpha\beta$ versus $\gamma\delta$ T-cell maturation \cite{seeland2018actin}. Similarly, Arhgdib, which encodes minor histocompatibility antigens (MiHA), has been implicated in antigen presentation and immune cell recognition \cite{pont2015lb}. A study has shown that T cells can recognize LB-ARHGDIB-1R (MiHA) on primary leukemic cells \cite{pont2015lb}. Together, the specific connections between gene group 17 and gene groups 9 and 39 highlight key processes involved in T-cell activation and antigen presentation during the early stages of glioblastoma progression. The absence of these interactions in late-GBM may indicate a shift in immune dynamics, potentially reflecting changes in the tumor microenvironment that alter T-cell function and immune surveillance. 

In the late-GBM network, gene group 17 forms a new connection with gene group 24 through edges between Apoe and Ftl1. Apoe encodes apolipoprotein E, a key component of plasma lipoproteins that not only regulates lipid metabolism but also induces tumor progression \cite{zhao2021apolipoprotein}. Notably, Apoe deficiency has been shown to support antitumor immunity by preventing T cell exhaustion, suggesting its role in immune suppression within the glioblastoma microenvironment \cite{zhao2021apolipoprotein}. Ftl1, encoding ferritin light polypeptide 1, is part of the ferritin complex, which plays a crucial role in iron storage and immune regulation \cite{wu2024ferritin}. Higher expression levels of ferritin components, including FTL and FTH, have been observed in regulatory T cells (Tregs) compared to conventional $CD4^+ CD127^+ CD25^-$ T cells, indicating a potential role in maintaining function of Tregs \cite{wu2024ferritin}. The presence of this connection in late-stage glioblastoma suggests a shift toward an immunosuppressive microenvironment, where connection between ApoE and FTL1 may contribute to tumor progression by promoting Treg-mediated immune evasion and T-cell exhaustion. This newly established interaction in late-GBM highlights a possible mechanism by which glioblastoma advances to a more immunosuppressive state, potentially limiting effective anti-tumor immune responses.

\section{Conclusions}

Joint gene network inference algorithms support the analysis of scRNAseq data by inferring and comparing networks between groups of cells. In this study, we proposed a new approach of joint graphical modeling using two-target linear shrinkage. TTLS shows better performance than individual network inference with GeneNet, and the fused graphical lasso, and has potential for large-scale analysis using less computational time compared to the Bayesian approach considered. Furthermore, higher criticism is highlighted as a second-level significance testing approach for use in high-dimensional partial correlation matrix estimation with high sparsity. Importantly, prior knowledge is not a requirement in JointStein which broadens application of JointStein in species with less available data.

In the analysis of experimental data from T cell clusters in a glioblastoma study, the potential of the JointStein framework in learning local network structures in heterogenous data is demonstrated. In future work we aim to extend this approach to heterogeneous samples with less clearly defined groups, where JointStein could be used to estimate local networks by sharing data from nearby samples. This will allow the approach to be applied to single cell data representing trajectories of cells, or to time series data.

\end{document}